\documentstyle[aps,prl,epsfig,twocolumn,floats,psfig]{revtex}
\advance\textheight by 0.14in
\advance\topmargin by +0.35in
\begin{document}
\twocolumn[\hsize\textwidth\columnwidth\hsize\csname @twocolumnfalse\endcsname
\title{Deconfinement and phase diagram of bosons in a linear optical lattice with a particle reservoir}
\author{Kingshuk Majumdar$^{(1)\!}$ and H.~A. Fertig$^{(2,3)\!}$}
\address{$^{(1)\!}$ Department of Physics, Berea College, Berea, KY 40404\\ 
$^{(2)\!}$ Department of Physics and Astronomy, University of Kentucky, Lexington, KY 40506-0055\\
$^{(3)\!}$Department of Physics, Indiana University, Bloomington, IN 47405}
\date{\today}
\maketitle

\begin{abstract}
{We investigate the zero-temperature phases of bosons in a one-dimensional optical lattice
with an explicit tunnel coupling to a Bose condensed particle reservoir. 
Renormalization group analysis of this system is shown to reveal three 
phases: one in which the
linear system is fully phase-locked to the reservoir; one in which Josephson vortices between
the one-dimensional system and the particle reservoir deconfine due to
quantum fluctuations, leading to a decoupled state in which the
one-dimensional system is metallic; and one in which the one-dimensional
system is in a Mott insulating state.}
\end{abstract}
\bigskip
]
\narrowtext
{\em Introduction:} Ultracold atoms trapped in optical lattices is an 
active area of research in many-body physics ~\cite{jaksch,stringari,pedri,pethick,moritz,shin}.
At zero temperature and without disorder, bosons in a 
lattice with integer filling exhibit two distinct
phases -- a superfluid phase, where the phase of the wavefunctions 
are sharply defined on sites, and 
a Mott insulating phase, where the site occupation numbers are sharply defined
~\cite{fisher,frey,fisher2,greiner,stoferley}. 
This system allows novel many-body phenomena associated with 
unusual correlated states and quantum
phase transitions~\cite{recati,buchler,kollath,demler,lieb,giamarchi,kuklov,zwerger}.

In this work we will study a system of bosons in a one-dimensional (1D) optical lattice 
that is tunnel coupled to a three-dimensional (3D) Bose-condensed particle reservoir. Our goal is to understand
how the reservoir impacts the states of the bosons in the optical lattice.
We find that the system supports three phases: a 3D superfluid phase 
in which the 1D system becomes phase-locked with the reservoir;
a decoupled phase (in which, under circumstances described
below, the 1D system behaves like a metal);
and a Mott insulating phase. These states are separated by deconfinement
transitions~\cite{fertig}, either of vortices or of tunneling events, as
we describe below. The states may be distinquished
both by their low-energy excitation spectra, and by their conduction 
properties.

{\em Model:} Our analysis begins with 
a Hamiltonian for lattice bosons (in number-phase representation)~\cite{fisher}
\begin{eqnarray}
{\cal H} &=& -t\sum_{\langle ij \rangle, \tau} \cos(\phi_{i\tau} - \phi_{j\tau}) + \frac {U}{2}\sum_{i,\tau} n_{i\tau}^2 \cr
&-&t_R\sum_{\langle {\bf R}{\bf R'} \rangle, \tau} \cos(\Phi_{{\bf R}\tau} - \Phi_{{\bf R'} \tau}) + \frac
{U_R}{2}\sum_{{\bf R},\tau} N_{{\bf R}\tau}^2 \cr 
&-&\sum_{i,\tau} J_i\cos (\phi_{i\tau} -\Phi_{{\bf R}(i) \tau}),
\label{ham}
\end{eqnarray}
where $(n_{i\tau},\phi_{i\tau})$ and $(N_{{\bf R}\tau}, \Phi_{{\bf R}\tau})$ are the canonically conjugate
occupation number fluctuations and phases of the bosons in the chain and 
in the 3D reservoir, respectively, $t$ and $U>0$ ($t_R$ and $U_R>0$)
are the nearest neighbor interchain hopping and on-site repulsion terms
for the 1D (3D) system,
and $J_i$ is a tunneling amplitude
between a site on the chain $[i]$ and a site in the 
reservoir $[{\bf R}(i)]$.  While the particle numbers in the real optical lattices are typically small, we 
nevertheless allow the fluctuations to vary from $-\infty$ to $+\infty$. This simplifies the calculation, and should not affect
the allowed phases because the symmetries of the resulting Hamiltonian are unchanged~\cite{cardy}. 
Our geometry could be realized as a linear array of sites
using red detuned light 
focused near the edge of a BEC cloud~\cite{recati,dumke,schlosser,kolomeisky}.
Note that the tight-binding form for
the reservoir is adopted purely as a matter of convenience.

Following standard procedure~\cite{negele}, we construct a path-integral 
representation of the partition function and 
then use the Villain model for the three cosine terms in 
Eq.~\ref{ham}~\cite{herbut}. 
Assuming that vortex 
rings in the 3D reservoir are unimportant (which will always be valid if the reservoir is sufficiently dilute), we integrate out 
the internal degrees of freedom of the reservoir to arrive at
a partition function of the form
\begin{eqnarray}
{\cal Z}_{VM}&=& 
\sum_{m_{x\tau},n_{x\tau}}\exp\Bigl\{-\frac 1{2} \sum_{x,\tau}
\Bigl[\varepsilon U |n_{x\tau}|^2 
+ \frac 1{\varepsilon t} |m_{x\tau}|^2 \Bigr] \cr
&-& \frac 1{2L_x\beta}\sum_{q_x,\omega_n} \frac 1{h({\bf q})}\Bigl|iq_xm({\bf q})+i\omega_n n({\bf q}) \Bigr|^2\Bigr \}.
\label{ZVM}
\end{eqnarray}
Here $\beta = (kT)^{-1} \rightarrow \infty$, 
$L_x$ is the number of sites in the 1D chain,
and $\varepsilon$  is the time slice interval. Physically,
the integer variables $m_{x\tau}$ may be understood as bond currents and 
$n_{x\tau}$ as the fluctuations in the site occupation number. Eq.~\ref{ZVM} is a form of the Bose-Hubbard model.
The coefficient $1/h({\bf q})$ 
contains information about the reservoir degrees of freedom, in particular the
gapless collective mode it supports due to its own superfluidity,
\begin{equation}
\frac 1{h({\bf q})} = \frac 1{\varepsilon J} + \gamma \ln \Bigl(1 + \frac{\Lambda^2 c_R^2}{c_R^2 q_x^2 + 
\omega_n^2}\Bigr), 
\label{heqn}
\end{equation}   
where $c_R = \sqrt{\varepsilon^2 t_RU_R}$, 
$\Lambda $ is the momentum cut-off, and $\gamma =(1/4\pi \varepsilon t_R)$ ~\cite{log}.

The partition function ${\cal Z}_{VM} \equiv \sum_{m,n} \exp(-{\cal H}_{VM})$
may be reexpressed
in terms of another pair of integer fields $\phi(x,\tau)$ and $A(x,\tau)$ with
$m(x,\tau) = -\partial_\tau \phi (x,\tau)$ and 
$n(x,\tau) = \partial_x \phi (x,\tau) + A(x,t)$, so that
\begin{eqnarray}
&&{\cal H}_{VM} = \frac 1{2K}\sum_{x,\tau} \Bigl\{c^2\Bigl|\partial_x \phi(x,\tau)+A(x,\tau)\Bigr|^2 \cr
&+& \Bigl|\partial_\tau \phi(x,\tau)\Bigr|^2 \Bigr\}
+ \frac 1{2L_x\beta}\sum_{q_x,\omega_n} \frac {1}{h({\bf q})}\Bigl|i\omega_n A({\bf q})\Bigr|^2.
\label{HamVM2}
\end{eqnarray} 
Here,
$K = \varepsilon t$ and $c=\sqrt{\varepsilon^2 Ut}$.  
Configurations for which $\vec{\nabla}\phi \ne 0$ and $A=0$ 
contain closed loops which may be understood as 
worldlines of particle-hole pairs that separate and recombine.  
The $A$ field, which should be regarded as residing on the 
time interval links, can cancel the gradient energy $(\partial_{x}\phi)^2$
on part of a closed loop configuration to 
form individual particle or
hole trajectories; the endpoints
(occuring where $\partial_{\tau}A(x,\tau) \ne 0$) represent tunneling events
between the 3D and 1D systems.   
Tunneling events
may be shown~\cite{fertig} to be dual to the vortices.
Moreover, the model defined by Eq.~\ref{HamVM2} may be generalized to
include a core energy $E_c$ for the vortices.

As we discuss below, this generalized Hamiltonian supports three phases.
When tunneling events proliferate through the system, the 1D and 3D systems exchange
particles freely (equivalently, vortices of the 1+1 dimensional
system are linearly confined),
and their phases become locked together to form a single superfluid.
As fluctuations in $\phi$ and $A$ 
are decreased (by decreasing $E_c$ and/or $h$, or decreasing $K$),
tunneling events bind into pairs which conserve the overall
particle number in the 1D system, although closed particle-hole worldlines
remain proliferated.
This state may be understood as one in which Josephson vortices 
between the 3D and 1D system proliferate due to quantum fluctuations,
effectively decoupling the two systems.
For still smaller fluctuations, particle-hole worldline loops of arbitrarily large size
become irrelevant,
indicating that the fluctuations
in the particle number on each site have been suppressed, and the system
is a Mott insulator.  

{\em RG analysis:} 
A method for constructing a momentum shell RG for Hamiltonians such 
as Eq.~\ref{HamVM2} was developed in Ref.~\onlinecite{fertig}.  We replace
the integer fields $\phi$ and $A$ with the
continuous fields $\varphi(x,\tau)$ and $a(x,\tau)$ in 
Eq.~\ref{HamVM2} and add terms
of the form $-y\int dxd\tau \cos[2\pi\varphi(x, \tau)]$ and 
$- y_a \int dxd\tau \cos[2\pi a(x,\tau)]$ so that the resulting effective Hamiltonian
has the same symmetries as the original one~\cite{cardy}.
In this replacement, $y=\exp(-E_c)$ is the usual vortex fugacity,
and choosing $y_a^2 \sim \int d^2q 1/h(q)$ approximately reproduces
the action associated with a worldline endpoint (i.e., a
tunneling event). We then integrate out short
wavelength degrees of freedom [$\Lambda/b <|q_x|,|w_n|/c < \Lambda$
with $b=\exp(l)$] to lowest order in $y$ and $y_a$, 
and rescale lengths, times, and fields according to 
$x=bx', \tau = b\tau', \varphi'(x',\tau') = \varphi(x,\tau)$,
and $a(x,\tau)=a'(x', \tau')/b$.  This choice preserves the
terms that are lowest order in gradients in the
quadratic part of the Hamiltonian. Because the $a$ field
shrinks upon rescaling, it is natural to expand $\cos[2\pi a(x,\tau)]$
in its argument, producing a quadratic term 
of the form $\frac 1{2}\rho |a(x,\tau)|^2$
that contributes
to the fixed point.  The higher order vertices generated by
this expansion contribute to the renormalization of $\rho$
but are themselves irrelevant. Note the initial value of $\rho$ is $4\pi^2 y_a$.

\begin{figure}[httb]
\protect \centerline{\epsfxsize=2.2in \epsfbox {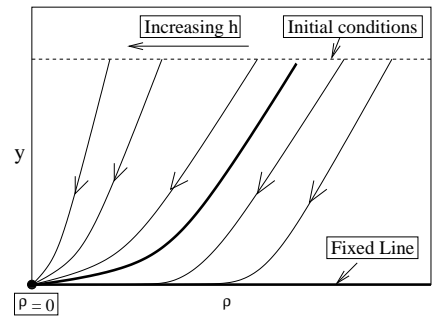}} \vskip 0.3cm 
\protect \caption{Schematic diagram of RG flows for the scaling relations
of $\rho$ and $y$ from initial microscopic values to a fixed line.  
$\rho=0$ represents a deconfined phase for
tunneling events.  For $\rho>0$ they are bound in pairs such that the
net particle flow from the reservoir vanishes.  Heavy line separates flows
for the two kinds of states.} 
\protect \label{phase}
\end{figure}

The fixed points that emerge from this procedure have
the form
\begin{eqnarray}
{\cal H}_{*} = \frac 1{2K}\int\!dx d\tau &\Bigl[& c^2\Bigl|\partial_x \varphi(x,\tau)+a(x,\tau)\Bigr|^2 \cr
&+& 
\Bigl|\partial_\tau \varphi(x,\tau)\Bigr|^2 
+ \rho K |a(x,\tau)|^2 \Bigr].
\label{Hfixed}
\end{eqnarray}  
The last term in ${\cal H}_{*}$
is very important: since
tunneling events are specified by $\partial_{\tau}a \ne 0$, when $\rho \ne 0$
they are bound into equal and opposite pairs.  Unbinding
occurs if $\rho(\ell)$ scales to zero.  To lowest
order in $y_a$, its scaling relation is
\begin{equation}
{{d \ln \rho} \over {d\ell}} = 
-\Bigl(\frac{2\pi \Lambda^2 {K}}{\sqrt{\rho K}}\Bigr)e^{-2\ell} .
\label{rhoflow}
\end{equation}
For small $\rho$ one can easily show that the term
$-y\int dxd\tau \cos[2\pi\varphi(x, \tau)]$ is strongly irrelevant.
The resulting scaling flows are shown in Fig.~\ref{phase}.
As may be seen, in general $\rho$ scales 
to a point along a fixed line; if its initial value is small enough
(as occurs for small $\gamma$ and large $J$), that point
is at $\rho=0$.  
A remarkable feature of Eq. \ref{rhoflow}
for such flows is that $\rho(\ell)=0$ at a {\it finite} value
of $\ell=\ell^*$, because of the singularity
as $\rho \rightarrow 0$.  If the initial value of $\rho$ is increased, $\ell^*$ eventually
diverges, defining a transition point above which $\rho$ scales
to a non-zero value. {\it This represents a confinement transition
for tunneling events}~\cite{fertig,com}. 

The dual representation of this system is obtained by
performing a Poisson resummation on $A$ and $\phi$ in
Eq.~\ref{HamVM2}. The resulting Hamiltonian is very similar in
form to Eq.~\ref{HamVM2}~\cite{nelsonbook}, and may be understood
as a model with vortex degrees of freedom rather than tunneling events.
Because of this, there is a second transition, in which vortices
go from a bound to an unbound state, in the same continuous 
fashion as seen above for tunneling events.  In terms of 
$\varphi$ and $a$ degrees of freedom, this corresponds to
a phase in which arbitrarily large particle-hole loops
are negligible in the partition function. Pairs of tunneling
events are then linearly confined~\cite{fertig}.

\begin{figure}[httb]
\protect \centerline{\epsfxsize=2.4in \epsfbox {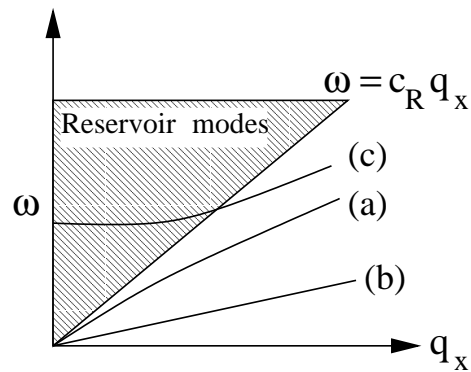}} \vskip 0.3cm 
\protect \caption{Collective mode spectra obtained from fixed point Hamiltonians. 
In the 3D superfluid phase there is a single sharp mode (a).  The 1D metallic
phase supports two gapless modes (a) and (b).  In the Mott insulator phase,
there is a only single gapped mode (c).  Note all three phases also contain a
continuum of modes due to the reservoir (shaded region).
}
\protect \label{mode}
\end{figure}

The phases 
described above have physically different characteristics.
This can be seen most directly in the collective mode spectra,
which in principle may be measured in light
scattering experiments~\cite{unpub,scattering}.
We find these by examining the 
density-density 
correlation function $\langle n(-q_x,-\omega)n(q_x,\omega) \rangle $
near the fixed point Hamiltonians representing the various
phases.
Generally, this contains a broad response for 
$\omega>c_R q_x$ due to modes in the reservoir.
Beyond this, in the superfuid phase [Fig. ~\ref{mode}(a)], 
we find~\cite{unpub} a sharp resonance (collective
mode) at $\omega_1 = c_R q_x - \delta \omega$, 
where $\delta \omega = c_R \Lambda \exp\{ -(c^2/\gamma K)/(q_x^2[c_R^2-c^2]) \}$.
This rapidly approaches the reservoir continuum
as $q_x \rightarrow 0$,
indicating that number fluctuations in the 1D system will strongly
mix with those of the reservoir in the long wavelength limit.
This is
consistent with our interpretation of this phase as a single 3D superfluid.
(Note that although the term containing the logarithmic singularity is irrelevant in the RG sense,
it nevertheless has the physical effect of 
``pulling'' the collective mode
very close to the edge of the reservoir modes 
at small enough wavevectors.)
In the intermediate phase we
find two collective modes, the superfluid mode of the reservoir at $\omega_1$ and 
another linear mode at $\omega_2 =
cq_x \sqrt{\rho K/(\rho K + c^2)}$ [Fig. ~\ref{mode}(b)]. 
For an appropriate geometry we will show this leads to {\it metallic} behavior.
In the Mott insulator phase, the 
reservoir becomes decoupled from the 1D chain in the
long-wavelength limit, and supports a
gapped mode at 
$\omega_3 = \sqrt{c^2q_x^2 + 2\pi^2 K/E_c}$ [Fig. ~\ref{mode}(c)]. 

\begin{figure}[ht]
\protect \centerline{\epsfxsize=3.0in \epsfbox {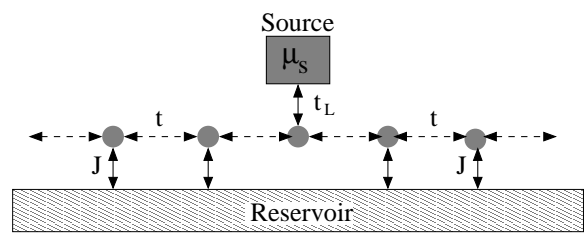}} \vskip .2cm 
\protect \caption{Josephson junction realization of the system with source
and drain to measure conductances of the phases.}
\protect \label{JJgeometry}
\end{figure}

{\em Conductance:} The different characters of the phases may also
be seen in the conductance~\cite{schultz} of the system.
To measure this, one has to 
attach a particle source and lead
as shown in Fig.~\ref{JJgeometry}. The conductance quantifies
the current injected by
this
source with chemical potential $\mu_s>0$,
draining into the 3D reservoir which is held at zero chemical potential.
(Note that a link has been removed to
ensure that the current flows in the 1D chain before 
tunneling into the reservoir.) This system could be
fabricated in a Josephson-junction array ~\cite{fisher}.

The source and lead introduce two extra degrees of freedom 
$[\phi_L(\tau), N_L(\tau)]$ to our model.
We treat the particle source as an ideal reservoir
with Hamiltonian $H_L=\mu_s N_L$;
current conservation at the site $(x=0)$ where 
current is injected specifies $N_L$ in terms of the
other variables through a
constraint in the partition function, 
$ \Pi_{\tau}\lbrace \partial_x m(x=0,\tau)+\partial_\tau n(x=0,\tau) 
- dN_L(\tau)/d\tau = 0 \rbrace$.
The conductance is given by $G(\omega) = 
\omega \langle N_L(-\omega) N_L(\omega) \rangle$.

The analysis proceeds in a fashion similar to what is 
described above for the uniform chain. However, because of the
lead there is an additional term in our effective Hamiltonian
of the form
$-\delta y \int d\tau \cos [2\pi a(x=0,\tau)]$,
reflecting the fact that the rate of tunneling events at the
$x=0$ site is different than at other sites. The scaling relation
for $\delta y(\ell)$ takes the form ~\cite{unpub} 
\begin{equation}
\frac {d \delta y (\ell)}{d\ell} \approx  \delta y (\ell) 
\Bigr[1 - \frac {\pi^2 K \Lambda^2 [\alpha (\ell)]^2}{\sqrt{\rho K}}\Bigl]. 
\label{yeqn1}
\end{equation}
for small $y$, $y_a$, and $\delta y$, with $\alpha(\ell)=\exp(-\ell)$.
It is important to recognize that in the 3D superfluid state,
$\rho(\ell) \rightarrow 0$ for {\it finite} $\ell \equiv \ell^*$,
so that $\delta y(\ell)$ will be driven to zero at $\ell=\ell^*$.
Thus the non-uniformity of tunneling events along the chain is
irrelevant in this state.  
In this situation, we can compute the conductance using
the fixed point Hamiltonian (Eq.~\ref{Hfixed}) for $\rho=0$
to find a true superconducting response, $G(\omega) \propto -i/(\omega+i\delta)$.

By contrast, if $\rho$ remains finite
as $\ell \rightarrow \infty$, $\delta y$ will necessarily grow,
and we need to look for a new fixed point.
To do this, we integrate the RG flows to a scale $\ell_0$ for which
the irrelevant operators may be ignored.  We then integrate out
$\varphi$ and $a(x \ne 0,\tau)$ and recollect some irrelevant
terms to restore the ${1 \over 2} \rho a(x=0,\tau)^2$ term to 
its cosine form $\cos [2\pi \alpha(\ell_0) a(x=0,\tau)]$, and
arrive at an effective Hamiltonian for the lead site in the
chain,
\begin{eqnarray}
{\cal H}_{eff}^L &=& 
\frac {c}{2K}\sqrt{\frac{\rho K}{\rho K + c^2}}\frac 1{\beta}\sum_{\omega_n}|\omega_n| |a(x=0,\omega_n)|^2 \cr
&+& y_0(\ell_0)\sum_{\tau}\cos \bigl[2\pi \alpha(\ell_0)a(x=0,\tau)\bigr].
\label{RGH}
\end{eqnarray}
We then modify the RG so that $a^{\prime}(x=0,\tau^{\prime})=a(x=0,\tau)$
to preserve the form of the quadratic term in ${\cal H}_{eff}^L$; the scaling relation
obeyed by $y_0$ is then
\begin{equation}
\frac {dy_0(\ell)}{d\ell} \approx  y_0 (\ell) 
\Bigr[1 - 2\pi^2 K \sqrt{\frac{\rho K + c^2}{\rho K c^2}}[\alpha (\ell_0)]^2\Bigl]. 
\label{yeqn2}
\end{equation}
It is important to recognize that $\alpha(\ell_0)$ does {\it not}
shrink as it did in Eq. \ref{yeqn1}. Thus
since $\rho$ is small, 
$y_0$ is irrelevant.  Our fixed point is then 
${\cal H}_{eff}^L $ with $y_0=0$.  In this case the conductance
is finite as $\omega \rightarrow 0$, with $G(\omega) \propto 1/\sqrt{\rho}$. This metallic behavior is surprising 
in light of the linear mode supported by this phase. It is a result of the very limited phase space available for
fluctuations in one dimension ~\cite{schultz}. For this reason the transport properties of this intermediate phase is
distinct from that of the 3D superfluid phase.

The conductance in the
deconfined vortex state can be computed straightforwardly using
the dual representation of the model.  The result unsurprisingly
is an insulating response $G(\omega) \sim \omega$.
(Details will be presented elsewhere~\cite{unpub}.)

In summary, we have shown that bosons in a one-dimensional optical
lattice that exchanges particles with a bulk superfluid supports
three distinct states, with different collective mode spectra
and conductances.

The authors thank K. MacAdam for helpful discussions.
Support was provided by the NSF under Grant Nos.~PHY9907949,
DMR0108451, and {DMR0454699}.
HAF thanks the KITP for its hospitality.

\end{document}